\documentclass[cameraready]{Interspeech}
\usepackage{amsmath, amssymb, amsfonts} % \mathbb, align, \eqref 

\usepackage[ruled,vlined]{algorithm2e}

\usepackage{graphicx}    % \includegraphics

\usepackage{booktabs}
\usepackage{multirow}
\usepackage{tabularx}
\usepackage{makecell}
\usepackage{siunitx}
\usepackage{subcaption} % 引入 subcaption 宏包
\usepackage{amsmath}
\usepackage{tikz}
\usetikzlibrary{shapes.geometric, arrows.meta, positioning, calc, shadows, backgrounds, fit, quotes}
\usepackage{fontawesome5} 
\title{Training-Free Multi-Step Inference for Target Speaker Extraction}

\author[affiliation={1}]{Zhenghai}{You}
\author[affiliation={2}]{Ying}{Shi}
\author[affiliation={1}]{Lantian}{Li}
\author[affiliation={2}]{Dong}{Wang}
\address{
    $^1$ Beijing University of Posts and Telecommunications, China \\
    $^2$ Center for Speech and Language Technologies, BNRist, Tsinghua University, China \
}

\email{youzhenghai@bupt.edu.cn, shiying@cslt.org, lilt@bupt.edu.cn, wangdong99@mails.tsinghua.edu.cn}


\usepackage{comment}

\begin{document}
\maketitle

\begin{abstract}

Target speaker extraction (TSE) aims to recover a target speaker’s speech from a mixture using a reference utterance as a cue. Most TSE systems adopt conditional auto-encoder architectures with one-step inference. Inspired by test-time scaling, we propose a training-free multi-step inference method that enables iterative refinement with a frozen pretrained model. At each step, new candidates are generated by interpolating the original mixture and the previous estimate, and the best candidate is selected for further refinement until convergence. Experiments show that, when ground-truth target speech is available, optimizing an intrusive metric (SI-SDRi) yields consistent gains across multiple evaluation metrics. Without ground truth, optimizing non-intrusive metrics (UTMOS or SpkSim) improves the corresponding metric but may hurt others. We therefore introduce joint metric optimization to balance these objectives, enabling controllable extraction preferences for practical deployment.

\end{abstract}

\section{Introduction}

In multi-speaker scenarios such as meetings, it is often necessary to isolate a target talker for downstream processing~\cite{chime8,yu2022m2met}. Target speaker extraction (TSE) addresses this problem by recovering the speech of a target speaker from a mixture given an enrollment utterance as a cue. Compared with generic source separation, which is agnostic to speaker identity, TSE is naturally aligned with target attribution and therefore avoids permutation ambiguity inherent in conventional speech separation tasks~\cite{zmolikova2023neural,elminshawi2022new}.

In recent years, end-to-end TSE systems have achieved substantial progress in model architectures, training strategies, and loss design~\cite{delcroix2020improving,xu2020spex, ge2020spex+}, leading to significant improvements in overall performance. Nevertheless, under challenging conditions such as highly similar speaker timbre, short enrollment utterances, or strong speaker overlap, the model may still suffer from target confusion or identity drift. In such cases, the extracted signal may gradually deviate from the target speaker or even collapse to the interfering speaker~\cite{zhao2022target}.

To mitigate these issues, most prior work focuses on improving the extraction and refinement capability of TSE models during training. For example, SpEx++~\cite{ge2021multi} and DPRNN-IRA~\cite{2020Robust} enhance robustness under difficult acoustic conditions through iterative extraction or reconstruction mechanisms, alleviating speaker confusion. X-SepFormer~\cite{liu2023x} reduces extraction errors and long-range identity drift through finer-grained chunk modeling, while X-TF-GridNet~\cite{hao2024x} and related architectures further improve separation quality and robustness using stronger backbones and improved training objectives. Although these approaches are highly effective, their gains typically rely on architectural redesign and retraining, and their inference-time enhancement capability remains tightly coupled with the training paradigm.

In contrast, a recent line of research improves output quality without modifying model parameters by allocating additional computation at inference time. In language modeling and reasoning, test-time scaling explores multiple candidate trajectories and improves final outputs through search, re-ranking, or iterative self-correction~\cite{yao2023tree,wei2022chain,ma2025thinking}. Similar training-free refinement strategies have recently shown promise in speech-related tasks. For example, in speech enhancement and source separation, test-time optimization or multi-step inference can provide additional gains without updating model parameters~\cite{zang2025training,raichle2026test}.

These approaches share a common intuition: a pretrained model already provides a strong inductive bias toward plausible solutions, while additional test-time computation is used to either explore alternative candidates or iteratively refine an initial estimate under a scoring signal. This process can reduce the gap between a single-pass prediction and a higher-quality solution without modifying model parameters.

Inspired by this perspective, we propose a training-free multi-step inference framework for TSE. At test time, we repeatedly reuse a frozen TSE model, generate a small set of candidate inputs by interpolating between the original mixture and the previous estimate, and select the best candidate using a deployable scoring function for the next refinement step. This procedure extends standard one-step TSE into an inference-time search process without requiring any parameter updates.

To better balance perceptual quality and target-speaker consistency in practical deployment, we further introduce a joint scoring function that combines non-intrusive quality prediction with speaker similarity. This joint objective guides the multi-step selection toward a more stable trade-off between perceptual quality and speaker consistency.

The contributions of this paper are two-fold:
(1) We propose a training-free multi-step inference framework for TSE that extends a one-step extractor into an inference-time search process via interpolation-based candidate construction and iterative selection using a frozen pretrained model. Oracle SI-SDRi selection on two representative TSE backbones demonstrates that the proposed search space can consistently improve performance over standard one-step inference without retraining.
(2) We further introduce a deployable joint scoring function based on UTMOS~\cite{saeki2022utmos} and speaker similarity (SpkSim) for non-intrusive candidate selection. The joint score achieves a more balanced improvement in perceptual quality and target-speaker consistency than single-metric selection.

% 需要在 Preamble 加上 \usepackage{fontawesome5} 来显示锁头图标
\definecolor{colorMixture}{RGB}{240, 240, 240}
\definecolor{colorModel}{RGB}{230, 240, 255}
\definecolor{colorSelector}{RGB}{255, 245, 230}
\definecolor{colorLine}{RGB}{80, 80, 80}

\begin{figure*}[tbh]
\centering
\begin{tikzpicture}[
    scale=0.85, % 稍微缩小比例
    transform shape,
    node distance=0.8cm and 1.0cm, 
    base/.style={draw, thick, align=center, minimum height=0.9cm, rounded corners=2pt, font=\small},
    input/.style={base, fill=gray!10, minimum width=1.2cm},
    proc/.style={base, fill=white, minimum width=2.0cm},
    model/.style={base, fill=blue!5, minimum width=2.4cm, minimum height=1.2cm, line width=1.2pt},
    select/.style={base, fill=orange!10, minimum width=2.2cm},
    arrow/.style={-Stealth, line width=0.8pt},
    dottedarrow/.style={-Stealth, line width=0.8pt, dashed, draw=gray!80}
]

    % --- 1. Inputs ---
    \node[input] (mixture) at (0,0) {Mixture $x_0$};
    \node[input] (enroll) [below=0.9cm of mixture] {Enroll $e$};

    % --- 2. Initial Inference ---
    \node[proc, right=0.8cm of mixture] (init) {Initial Inference\\$\hat{s}_0 = f_{\theta}(x_0, e)$};

    % --- 3. Iterative Loop Components ---
    \node[proc, right=1.0cm of init] (interp) {Interpolation\\Eq. (2)};
    
    % 模型叠加效果 (Parallel K)
    \node[model, right=1.0cm of interp, xshift=5pt, yshift=5pt] (m3) {};
    \node[model, right=1.0cm of interp, xshift=2.5pt, yshift=2.5pt] (m2) {};
    \node[model, right=1.0cm of interp, fill=blue!5] (m1) {Frozen TSE\\Model $f_{\theta}$ \ \faLock}; 

    \node[select, right=1.0cm of m1] (scoring) {Selector $R(\cdot)$};
    \node[circle, draw, thick, right=0.5cm of scoring, inner sep=1pt, font=\scriptsize] (argmax) {$\arg\max$};
    
    % 中间迭代输出 (t+1)
    \node[right=0.6cm of argmax, font=\small] (st) {$\hat{s}_{t+1}$};
    
    % 最终输出 (T) 
    \node[right=0.8cm of st, font=\large\bfseries] (sT) {$\hat{s}_{T}$};

    %  连线
    \draw[arrow] (mixture) -- (init);
    \draw[arrow] (init) -- (interp);
    \draw[arrow] (interp) -- (m1);
    \draw[arrow] (m1) -- (scoring);
    \draw[arrow] (scoring) -- (argmax);
    \draw[arrow] (argmax) -- (st);
    \draw[arrow] (st) -- node[above, font=\scriptsize] {$t=T$} (sT);

    % Mixture
    \draw[arrow] (mixture) |- ($(interp.north) + (0, 0.4)$) -- (interp.north);
  
    % Enrollment
    \draw[arrow] (enroll.east) -- (enroll.east -| init.south) -- (init.south);
    \draw[arrow] (enroll.east) -- (enroll.east -| m1.south) -- (m1.south);
    \draw[arrow] (enroll.east) -- (enroll.east -| scoring.south) -- (scoring.south);  

    % --- 反馈回路
    \draw[dottedarrow] (st.south) -- ++(0,-0.8) -| node[pos=0.25, fill=white, font=\scriptsize, inner sep=2pt] {Update estimate for $t+1$} (interp.south);
  
    % --- 装饰性标签与背景框 ---
    \begin{scope}[on background layer]
        % 【关键修改】：fit 增加了 (st)，使阴影覆盖 s_{t+1}
        \node[fill=blue!5, draw=blue!20, dashed, inner sep=8pt, rounded corners=5pt, fit=(interp) (m3) (st)] (loop_box) {};
        
        \node[anchor=south west, font=\footnotesize\bfseries\itshape, color=blue!60!black] at (loop_box.north west) {Inference-time Search (Iterated $T$ times)};
        
        \node[anchor=south, font=\scriptsize\color{blue!80!black}] at (m3.north) {$K$ Candidates};
    \end{scope}

\end{tikzpicture}
\caption{Overall architecture of the proposed multi-step inference TSE. The frozen model $f_\theta$ is shared across iterations to progressively refine the target estimate via input interpolation and non-intrusive joint selection.}
\label{fig:method_overview}
\end{figure*}
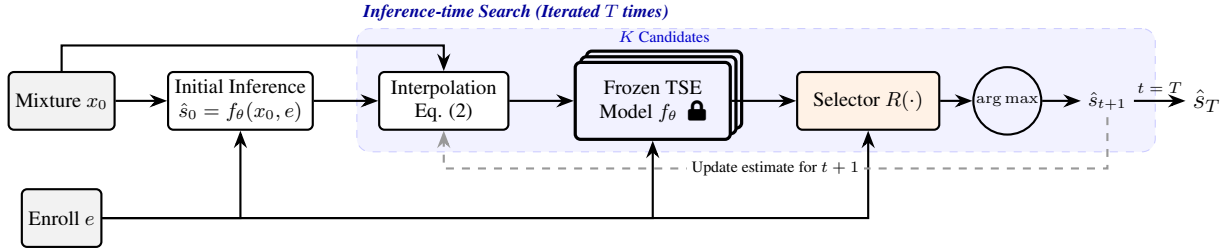

\section{Related work}
\label{sec:work}

The work most closely related to ours is a recent training-free multi-step inference approach for source separation~\cite{zang2025training}. By searching over remix-based candidates using a frozen pretrained model, that work demonstrates clear and reproducible inference-time gains beyond a single forward pass, showing that non-trivial performance headroom can be unlocked without retraining. In their setting, the most reliable improvements are obtained when candidate selection is guided by intrusive metrics such as PESQ or SDR. However, the results also highlight the issue of metric bias: optimizing one selection metric can degrade performance under other evaluation objectives.

Directly transferring this idea to TSE, however, is non-trivial. In addition to separation quality and perceptual fidelity, TSE must also preserve consistency with an enrolled target speaker. Moreover, in real-world deployment, clean target references are unavailable, making intrusive selectors unsuitable for online selection. These constraints call for a deployable scoring mechanism that explicitly balances non-intrusive quality assessment with enrollment-based speaker similarity, which is the focus of our method.

\section{Method}

\subsection{Multi-step candidate search via input interpolation}
\label{subsec:multistep_search}

Given a single-channel mixture $x_0\in\mathbb{R}^{L}$ and an enrollment utterance of the target speaker $e\in\mathbb{R}^{L_e}$, TSE aims to recover the target speech $s\in\mathbb{R}^{L}$.
Let $f_{\theta}$ denote a pretrained TSE model with frozen parameters. The standard one-step inference is:
\begin{equation}
\hat{s}_0 = f_{\theta}(x_0, e).
\label{eq:one_step_tse}
\end{equation}

Without updating model parameters, we improve the output via inference-time candidate search. Specifically, at iteration $t$ ($t=1,2,\dots,T$), we construct candidate inputs by linearly interpolating the previous estimate $\hat{s}_{t-1}$ and the original mixture $x_0$.
Let $\{r_t^{(k)}\}_{k=1}^{K}$ be $K$ interpolation coefficients with $r_t^{(k)}\in[0,1]$, and we explicitly include the endpoint $r_t^{(1)}=1$ (optionally also $r_t^{(K)}=0$). The $k$-th candidate input is
\begin{equation}
x_t^{(k)} = r_t^{(k)} x_0 + \bigl(1 - r_t^{(k)}\bigr)\hat{s}_{t-1}, \qquad k=1,\dots,K
\label{eq:candidate_input}
\end{equation}
and the corresponding candidate output is obtained by reusing the same frozen model:
\begin{equation}
\hat{s}_t^{(k)} = f_{\theta}\bigl(x_t^{(k)}, e\bigr).
% , \qquad k=1,\dots,K,
\label{eq:candidate_output}
\end{equation}

Given an inference-time scoring function $R(\cdot)$, we perform greedy selection at each step and take the highest-scoring candidate as the next estimate:
\begin{equation}
k_t^\ast = \arg\max_{k} R\bigl(\hat{s}_t^{(k)}; e\bigr), \qquad
\hat{s}_t = \hat{s}_t^{(k_t^\ast)}.
\label{eq:select_best}
\end{equation}
% The scoring function $R(\cdot)$ specifies the selection criterion of inference-time search.
% The selector $R(\hat{s};e)$ specifies the optimization target of inference-time search and can incorporate any deployable signals available at test time. 

We use SI-SDRi as an oracle selector to estimate the upper bound of the candidate space. In deployable settings , 
we consider a non-intrusive quality predictor, UTMOS~\cite{saeki2022utmos}, and an enrollment-based speaker similarity score (SpkSim). To balance perceptual quality and target-speaker consistency, we propose a joint selector defined as:
\begin{align}
R_{\mathrm{joint}}(\hat{s}; e)
&= \mathrm{UTMOS}(\hat{s}) \nonumber\\
&\quad + \lambda \Bigl(1-\exp\bigl(-\alpha \cdot \mathrm{SpkSim}(\hat{s}, e)\bigr)\Bigr).
\label{eq:joint_score_sat}
\end{align}
where $\lambda$ balances the relative scales of the two terms, and $\alpha$ controls the saturation in the high-similarity region to avoid over-emphasizing identity consistency in later iterations.

\subsection{Reliability analysis}
\label{subsec:theory}

Our multi-step inference introduces an inference-time search loop driven by a scoring function $R$, which raises two reliability questions in practice. We ask whether greedy selection provides a safe fallback, i.e., whether the search can ever be worse than the one-step output under the optimized selector, and how sensitive the search trajectory is to imperfect scoring function, especially when $R$ is deployable but biased or noisy. We address these questions with two complementary analyses: the non-decreasing property under greedy selection and an error-bound analysis that bounds the score deviation caused by imperfect selector.

The proof strategy is inspired by recent training-free multi-step inference analysis in audio separation~\cite{zang2025training}, but we restate it here in the conditional TSE setting, where the extractor is conditioned on an enrollment utterance and the selector can incorporate enrollment-dependent signals. 
With the enrollment $e$ and the mixture $x_0$ fixed, we write the conditional mapping as $f_{\theta,e}(\cdot):\mathbb{R}^L\rightarrow\mathbb{R}^L$ and the scoring function as $R(\cdot; e)$. For brevity, we denote them by $f(\cdot)$ and $R(\cdot)$, respectively.

\textbf{(1) Non-decreasing property.}
Since the candidate set explicitly includes $r_t^{(1)}=1$, the corresponding candidate input is $x_t^{(1)}=x_0$, and the candidate output satisfies
$\hat{s}_t^{(1)} = f(x_0,e) = \hat{s}_0$.
Because Eq.~(\ref{eq:select_best}) selects the highest-scoring candidate at each iteration, we have
\begin{equation}
R(\hat{s}_t)
= \max_{k} R\bigl(\hat{s}_t^{(k)}\bigr)
\ge R\bigl(\hat{s}_t^{(1)}\bigr)
= R(\hat{s}_0).
\label{eq:nondecreasing_tse}
\end{equation}
Therefore, the multi-step search is guaranteed to be non-worse than the initial one-step output under metric $R$, and there always exists a fallback path to $\hat{s}_0$.

\textbf{(2) Error bound.}
Because the scoring function may be imperfect, the selected interpolation coefficient $\tilde r_t$ can deviate from the optimal $r_t^{\star}$.
Let $x_t^{\star}$ and $\tilde x_t$ be the optimal and selected inputs, respectively, and let $s_t^{\star}=f(x_t^{\star})$ and $\tilde s_t=f(\tilde x_t)$ be the corresponding outputs. From Eq.~(\ref{eq:candidate_input}), the input deviation is
\begin{equation}
\Vert \tilde x_t-x_t^{\star}\Vert _2
= |\tilde r_t-r_t^{\star}| \cdot \Vert x_0-\hat{s}_{t-1}\Vert _2.
\label{eq:xt_diff}
\end{equation}

We assume that $f$ is Lipschitz continuous w.r.t.\ its input, and $R$ is Lipschitz continuous w.r.t.\ the waveform input: for any two inputs, there exist constants $L_f, L_R>0$ such that
\begin{align}
&|R(a)-R(b)| \le L_R \Vert a-b\Vert _2, \nonumber\\
&\Vert f(u)-f(v)\Vert _2 \le L_f \Vert u-v\Vert _2.
\label{eq:lipschitz_assump}
\end{align}

Combining Eqs.~(\ref{eq:xt_diff})--(\ref{eq:lipschitz_assump}), we obtain the following upper bound on the score deviation:
\begin{align}
\bigl|R(\tilde s_t)-R(s_t^{\star})\bigr|
& \le L_R L_f\, |\tilde r_t-r_t^{\star}| \cdot \Vert x_0-\hat{s}_{t-1}\Vert _2 .
\label{eq:error_bound_det}
\end{align}

If we further model the ratio error $\Delta r_t=\tilde r_t-r_t^{\star}$ as a zero-mean random perturbation with $\mathrm{Var}(\Delta r_t)=\varepsilon_r^2$, then
\begin{align}
 \mathrm{Var}\bigl(R(\tilde s_t)\bigr) 
 \le (L_R L_f)^2 \, \Vert x_0-\hat{s}_{t-1}\Vert _2^2 \, \varepsilon_r^2 .
\label{eq:error_bound_var}
\end{align}

Eqs.~(\ref{eq:error_bound_det})--(\ref{eq:error_bound_var}) indicate that the impact of imperfect scoring on the optimized metric is quantifiable: its sensitivity scales with the interpolation-segment length $\Vert x_0-\hat{s}_{t-1}\Vert _2$ and is jointly controlled by the local smoothness of the model and the score (through $L_f$ and $L_R$). For well-trained models with more stable local behavior, the effective Lipschitz constants are typically smaller or better controlled ~\cite{gouk2021regularisation,finlay2018lipschitz,miyato2018spectral}, which helps keep the bound tight even with noisy scores. Moreover, as the search process stabilizes and the segment length becomes smaller or less variable, the bound correspondingly tightens, providing an interpretable stability risk bound for multi-step candidate search.

Overall, the non-decreasing property provides a safe fallback with respect to the optimized selector, while the error-bound analysis offers an interpretable stability intuition on how proxy imperfections affect the search trajectory.

\section{Experiments}
\label{sec:experiments}

\subsection{Experimental setup}

\subsubsection{Data}
\label{subsubsec:data}

We evaluate the proposed method on Libri2Mix~\cite{cosentino2020librimix}. 
Libri2Mix is a two-speaker, single-channel mixture dataset built from LibriSpeech~\cite{panayotov2015librispeech}. 
Models are trained on the 16\,kHz \emph{train-100} subset. 
The mixture SNR has a mean of 0\,dB and a standard deviation of 3.6\,dB, covering a wide range of mixture difficulty levels and making it a suitable testbed for constructing interpolation-based multi-step refinement trajectories. 
Each sample includes the mixture, the clean target speech, and an additional enrollment utterance from the same target speaker.

\subsubsection{Model training setup}
\label{subsubsec:train_setup}

To verify that our method generalizes across TSE architectures, we adopt two representative end-to-end backbones: a DPRNN-based TSE system~\cite{2020Dual} and a lighter SpEx+ baseline~\cite{ge2020spex+}. 
Their single-pass MACs are 121.646\,G and 7.425\,G, respectively.
Both backbones are trained following the configuration in~\cite{you2025spkaugtse}. The speaker encoder consists of three ResNet blocks and produces a 128-dimensional speaker embedding. 
During training, we jointly optimize an SI-SDR loss~\cite{luo2018tasnet} and a cross-entropy speaker classification loss for speaker discrimination. 

During training, mixture and target signals are cropped into fixed 4\,s segments, and the enrollment utterance is randomly sampled from another utterance of the same target speaker. All models are trained with Adam for up to 200 epochs with an initial learning rate of 0.001. The learning rate is halved if the validation loss does not improve for two consecutive epochs. The source code is publicly available.\footnote{\url{https://github.com/youzhenghai/mutli-step-TSE}}

\subsubsection{Multi-step inference and evaluation setup}
\label{subsubsec:infer_eval_setup}

Our method is strictly training-free: at test time, all TSE model parameters are frozen, and only the input construction and candidate selection strategy are extended. Unless otherwise specified, we perform $T=5$ refinement steps and sample $K=20$ interpolation candidates per step, with interpolation coefficients uniformly sampled from $[0,1]$ to construct candidate inputs.
The computational cost grows approximately linearly with the number of evaluated candidates, making the method suitable for post-training enhancement scenarios where additional inference computation is acceptable.

We evaluate three aspects of TSE outputs: separation quality, perceptual quality, and target-speaker consistency, measured by SI-SDRi~\cite{le2019sdr}, UTMOS\footnote{\url{https://github.com/sarulab-speech/UTMOS22}}, and SpkSim (cosine similarity between the estimate and the enrollment utterance in the embedding space of a pretrained CAM++~\cite{wang2023cam++} speaker encoder), respectively. 
For candidate selection, SI-SDRi serves as an oracle selector for upper-bound estimation, while UTMOS, SpkSim, and the proposed joint score in Eq.~(\ref{eq:joint_score_sat}) are used as deployable selectors; 
we set $\lambda=2.5$ and $\alpha=4.0$ for all experiments based on validation statistics. Table~\ref{tab:unified_results_main} reports results at Steps~1, 3, and 5; the omitted intermediate steps follow the same trend.

\begin{table*}[th]
\centering
\caption{Multi-step inference results on Libri2Mix-clean test under different candidate selectors. Step~0 denotes the common baseline for all methods. The ``Joint'' selector (UTMOS + SpkSim) follows the scoring $R(\cdot)$ in Eq.~(\ref{eq:joint_score_sat}).}
\label{tab:unified_results_main}
\setlength{\tabcolsep}{5pt} % 稍微增加列间距提高可读性
\renewcommand{\arraystretch}{1.0} % 恢复标准行高，避免过于拥挤
\resizebox{0.72\textwidth}{!}{
\begin{tabular}{ll ccc ccc}
\toprule
\multirow{2}{*}{\textbf{Selectors}} & \multirow{2}{*}{\textbf{Step}}
& \multicolumn{3}{c}{\textbf{DPRNN}}
& \multicolumn{3}{c}{\textbf{SpEx+}} \\
\cmidrule(lr){3-5} \cmidrule(lr){6-8}
& & SI-SDRi$\uparrow$ & UTMOS$\uparrow$ & SpkSim$\uparrow$
  & SI-SDRi$\uparrow$ & UTMOS$\uparrow$ & SpkSim$\uparrow$ \\
\midrule

% 提取出来的公共 Baseline
Baseline & 0 & 14.422 & 3.058 & 0.671 & 13.729 & 2.863 & 0.629 \\
\midrule 
% \midrule % 双线区分 Baseline 与后续 Selection 方法

\multirow{3}{*}{SI-SDRi (oracle)}
& 1 & \textbf{15.369} & 3.107 & 0.672 & 14.380 & 2.935 & 0.633 \\
& 3 & 15.241 & 3.107 & 0.671 & 14.387 & 2.932 & 0.632 \\
& 5 & 15.241 & 3.111 & 0.672 & \textbf{14.404} & 2.931 & 0.631 \\
\cmidrule(lr){1-8} % 细分割线

\multirow{3}{*}{UTMOS}
& 1 & 14.287 & 3.206 & 0.673 & 13.693 & 3.019 & 0.629 \\
& 3 & 14.037 & 3.242 & 0.674 & 13.596 & \textbf{3.036} & 0.626 \\
& 5 & 13.904 & \textbf{3.246} & 0.674 & 13.536 & 3.033 & 0.624 \\
\cmidrule(lr){1-8}

\multirow{3}{*}{SpkSim}
& 1 & 13.845 & 3.064 & 0.692 & 13.897 & 2.867 & 0.649 \\
& 3 & 13.215 & 3.057 & 0.697 & 13.701 & 2.849 & 0.652 \\
& 5 & 12.897 & 3.049 & \textbf{0.698} & 13.627 & 2.839 & \textbf{0.652} \\
\cmidrule(lr){1-8}

\multirow{3}{*}{Joint}
& 1 & 14.311 & 3.204 & 0.677 & 13.876 & 3.013 & 0.634 \\
& 3 & 14.181 & 3.238 & 0.679 & 13.772 & \textbf{3.028} & 0.634 \\
& 5 & 14.144 & \textbf{3.242} & \textbf{0.679} & 13.728 & 3.025 & \textbf{0.634} \\
\bottomrule
\end{tabular}
}
\end{table*}

\subsection{Experimental results}
\label{sec:results}

\subsubsection{Oracle upper-bound search}

Using SI-SDRi as an oracle selector aligns the search objective with offline evaluation, and therefore serves as an upper-bound probe of the proposed candidate space. Table~\ref{tab:unified_results_main} shows that both backbones expose clear headroom beyond Step~0 under this oracle setting.

For DPRNN, the main gain appears immediately: the best SI-SDRi is reached at Step~1 (15.369\,dB, +0.947\,dB over Step~0), after which the curve fluctuates mildly but stays above the one-step baseline. SpEx+ behaves differently: its best SI-SDRi is attained at a deeper step (14.404\,dB at Step~5, +0.675\,dB), suggesting that the lighter backbone benefits from a longer correction trajectory. This difference in best-step depth is consistent with the intuition that multi-step inference can adapt to backbone-specific error-correction dynamics.

Beyond SI-SDRi, oracle selection also coincides with small improvements in deployable metrics at some steps, indicating that the interpolation-based candidates can sometimes move waveform fidelity, perceptual quality, and speaker consistency in the same direction. Overall, the oracle results support the core premise of this work: inference-time search over the mixture--estimate bridge contains non-trivial performance headroom for TSE without retraining.

\subsubsection{Non-intrusive single-metric search}

In deployment, clean target references are typically unavailable, so candidate selection must rely on deployable scoring signals. The UTMOS-only and SpkSim-only settings in Table~\ref{tab:unified_results_main} exhibit a consistent pattern across both backbones: the search reliably pushes the optimized score upward, while the unoptimized objectives may drift.

With UTMOS as the selector, perceptual quality improves and then gradually saturates (e.g., DPRNN reaches 3.246 at Step~5; SpEx+ reaches 3.036 at Step~3). With SpkSim as the selector, enrollment-based similarity strengthens steadily (up to 0.698 for DPRNN and 0.652 for SpEx+ at Step~5). These trends confirm that the candidate space contains deployable improvements along both axes, and that each proxy can effectively steer the search toward its own target.

A more subtle observation is that the optimized score is not strictly monotonic step by step, yet it remains above the Step~0 baseline at all reported steps and tends to level off in later iterations. This matches our analysis: the non-decreasing property in Eq.~(\ref{eq:nondecreasing_tse}) guarantees a Step~0 fallback under greedy selection, while the error-bound results in Eqs.~(\ref{eq:error_bound_det})--(\ref{eq:error_bound_var}) suggest that as refinement stabilizes, the effective interpolation distance $\Vert x_0-\hat{s}_{t-1}\Vert_2$ shrinks, making later updates smaller and the trajectory less sensitive to proxy noise.

At the same time, single-metric bias is evident. SpkSim-driven selection improves target-speaker consistency but often trades off SI-SDRi, whereas UTMOS-driven selection yields clearer perceptual gains with limited and less stable improvements in speaker similarity. This mismatch highlights the limitation of optimizing a single proxy in a multi-objective setting and motivates joint scoring for a more controlled trade-off.

\subsubsection{Joint-metric search}

We next evaluate the proposed joint selector in Eq.~(\ref{eq:joint_score_sat}), which combines non-intrusive quality prediction with enrollment-based speaker similarity. As shown in Table~\ref{tab:unified_results_main}, joint scoring improves both deployment-oriented metrics on both backbones (DPRNN: UTMOS 3.242 and SpkSim 0.679 at Step~5; SpEx+: UTMOS 3.028 and SpkSim 0.634 at Step~5), providing a more balanced refinement direction than either single proxy alone.

Meanwhile, SI-SDRi under joint scoring is not monotonic with step depth and can decline at deeper steps. This is expected because SI-SDRi is a waveform-level fidelity metric, and candidates favored by perceptual predictors or similarity scores are not necessarily those that minimize waveform reconstruction error. In addition, non-intrusive quality predictors inevitably carry domain shift and estimation noise, further weakening alignment between inference-time selectors and reference-based metrics.

Taken together, oracle selection defines the headroom of the candidate space, while deployable selectors determine how that headroom can be accessed in practice. Joint scoring offers a practical compromise by improving perceptual quality and target-speaker consistency without clean references, and provides a more controlled refinement direction than single-metric selection. While there is still room to better approach the oracle upper bound, these results suggest that training-free multi-step inference is a promising post-training enhancement option for deployment-oriented TSE.

\section{Conclusion}
\label{sec:conclusion}

We presented a training-free multi-step inference framework for target speaker extraction. Using a frozen pretrained TSE model, we construct interpolation-based candidates between the mixture and the current estimate and perform score-guided selection for iterative refinement, transforming standard one-step inference into an inference-time search procedure.
Experiments on two representative TSE backbones show that oracle SI-SDRi selection exposes clear headroom beyond one-step inference. Deployable non-intrusive selectors (UTMOS and SpkSim) further confirm the practical value of inference-time search, while also revealing the bias of single-metric optimization. In contrast, the proposed joint scoring achieves a more stable balance between perceptual quality and target-speaker consistency without reference signals.

These results suggest that multi-step inference is a practical post-training enhancement direction for TSE, especially when deployment requires improving quality and target consistency without retraining. Future work will explore more reliable non-intrusive scoring and calibration that better capture target confusion, with the goal of narrowing the gap between deployable selection and oracle upper-bound performance.

\section{Generative AI Use Disclosure}
Generative AI tools were used only for grammar correction of parts of the manuscript. The authors reviewed and revised all AI-assisted edits and take full responsibility for the content of the paper.

\bibliographystyle{IEEEtran}
\bibliography{mybib}

\end{document}